\documentclass[runningheads]{llncs}
\usepackage{graphicx}
\usepackage{comment}
\usepackage{amsmath,amssymb} 
\usepackage{color}
\usepackage{algorithm, algorithmic}
\usepackage{booktabs}
\usepackage{multirow}
\usepackage{bm}
\newcommand{\tabincell}[2]{\begin{tabular}{@{}#1@{}}#2\end{tabular}}

\begin{document}
\pagestyle{headings}
\mainmatter
\def\ECCVSubNumber{5130}  

\title{PAMS: Quantized Super-Resolution via Parameterized Max Scale}

\titlerunning{PAMS: Quantized Super-Resolution via Parameterized Max Scale}
%
\author{Huixia Li\inst{1 \dag}, Chenqian Yan\inst{1 \dag},
Shaohui Lin\inst{2},
Xiawu Zheng\inst{1}, \\
Yuchao Li\inst{1}, Baochang Zhang\inst{3},
Fan Yang\inst{4},
Rongrong Ji\inst{15}\thanks{Corresponding author.}
}
 \authorrunning{Huixia Li et al.}
%
\institute{$^{1}$Media Analytics and Computing Lab, Department of Artificial Intelligence, \\ School of Informatics, Xiamen University, China. \\
$^{2}$
National University of Singapore, Singapore.  $^{3}$ Beihang University, China.  \\
$^{4}$
Huawei Technologies Co., Ltd.  $^{5}$Peng Cheng Laboratory. \\
\email{\{hxlee, zhengxiawu\}@stu.xmu.edu.cn,} \\ 
\email{\{im.cqyan, shaohuilin007, xiamenlyc\}@gmail.com,}\\
\email{bczhang@buaa.edu.cn, yangfan74@huawei.com, rrji@xmu.edu.cn}
}

\renewcommand{\thefootnote}{\fnsymbol{footnote}}
\footnotetext[4]{Equal contribution. }

\maketitle

\begin{abstract}
Deep convolutional neural networks (DCNNs) have shown dominant performance in the task of super-resolution (SR). However, their heavy memory cost and computation overhead significantly restrict their practical deployments on resource-limited devices, which mainly arise from the floating-point storage and operations between weights and activations. Although previous endeavors mainly resort to fixed-point operations, quantizing both weights and activations with fixed coding lengths may cause significant performance drop, especially on low bits. Specifically, most state-of-the-art SR models without batch normalization have a large dynamic quantization range, which also serves as another cause of performance drop. To address these two issues, we propose a new quantization scheme termed PArameterized Max Scale (PAMS), which applies the trainable truncated parameter to explore the upper bound of the quantization range adaptively. Finally, a structured knowledge transfer (SKT) loss is introduced to fine-tune the quantized network. Extensive experiments demonstrate that the proposed PAMS scheme can well compress and accelerate the existing SR models such as EDSR and RDN. Notably, 8-bit PAMS-EDSR improves PSNR on Set5 benchmark from 32.095dB to 32.124dB with 2.42$\times$ compression ratio, which achieves a new state-of-the-art.

\end{abstract}

	\begin{figure*}[ht]

		\centering
		\setlength{\abovecaptionskip}{0cm}
	    \setlength{\belowcaptionskip}{0cm}
		\includegraphics[width=0.99\textwidth, height=0.39\textwidth]{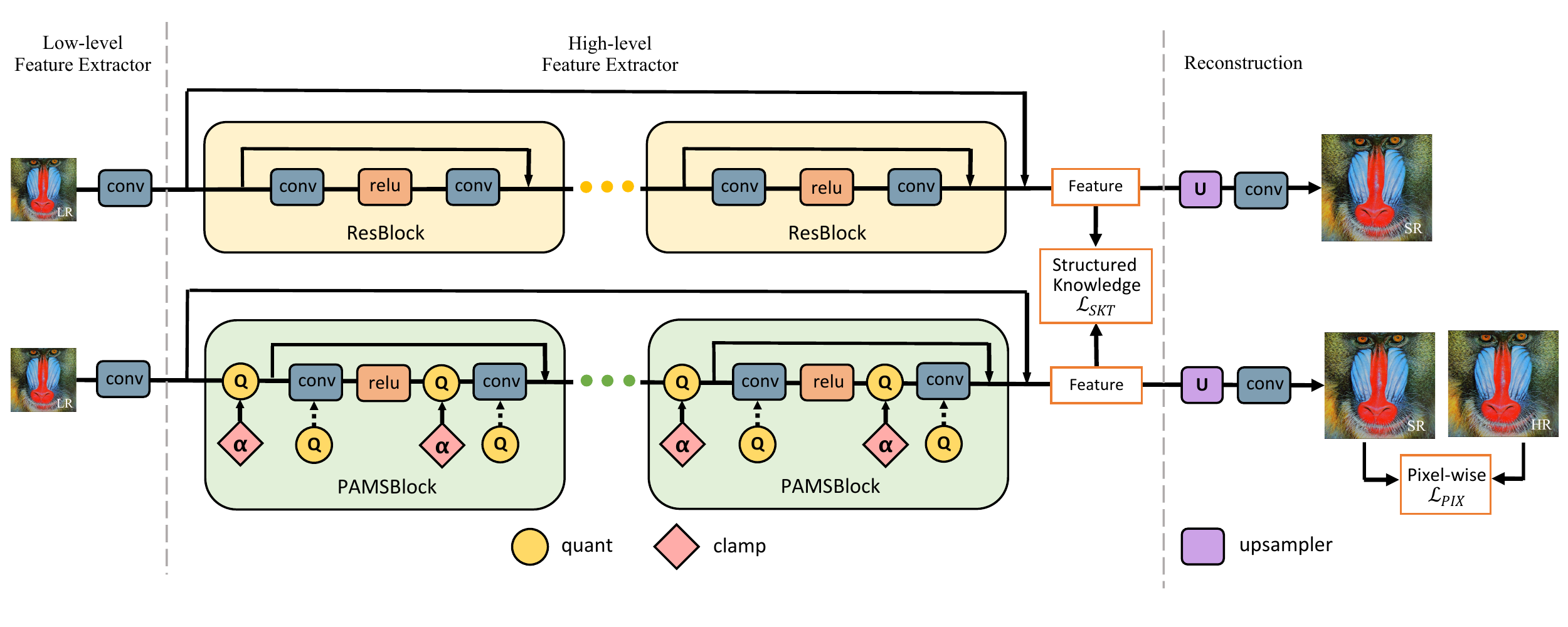} 
		\caption{The framework of our approach. The super-resolution operation is split into three modules, \emph{i.e.}, low-level feature extractor, high-level feature extractor and reconstruction. We deploy PAMS with different $\alpha$ on each activation layer in the high-level feature extractor. $quant$ denotes the quantization operation and $clamp$ represents the clamp function of quantization. Dash lines denote the weights are quantized with the maximum. Here, we illustrate EDSR as backbone.}
		\label{framework}
	\end{figure*}
	
\section{Introduction}
Single image super-resolution (SISR) aims to recover a high-resolution (HR) image from the corresponding low-resolution (LR) one, which has been a research hot spot in computer vision for decades. Coming with the advances of deep learning, deep convolutional neural networks (DCNNs)  \cite{dong2014learning,kim2016accurate,lim2017enhanced} have dominated SR in recent years. These networks commonly use an extraction module to extract a series of feature maps from the LR image, cascaded with the up-sampling module, which stepwisely increases the resolution to reconstruct the HR image.

As one of the pioneering works for deep learning based SR, Dong \emph{et al.} \cite{dong2014learning} introduce three convolution layers to achieve high visual perception. After that, Kim \emph{et al.} \cite{kim2016accurate} design a deep network VDSR by stacking 20 convolutional layers. Subsequent works mainly resort to increasing the network depth to improve SR performance. For instance, Lim \emph{et al.} \cite{lim2017enhanced} propose the enhanced deep residual networks (\emph{e.g.} EDSR and MDSR) and remove batch normalization (BN) \cite{ioffe2015batch} to reduce the memory consumption, which however still requires at least 64 convolution layers (more than 160 layers for MDSR). A channel attention mechanism equipped into the RCAN model \cite{zhang2018image} requires more than 400 layers with about 30B FLOPs and 13M parameters. Such significant computation and memory overheads restrict their applications in scenarios where only limited memory and computation resources are available. Consequently, compressing deep SR networks has attracted increasing attention recently \cite{ma2018efficient}. 

Beyond SR, neural network compression and acceleration have been widely studied in the literature. Representative works include parameter pruning \cite{He_2017_ICCV,he2018soft,he2019filter,lin2018accelerating,lin2019towards,li2019exploiting}, low-rank approximation \cite{denton2014exploiting,lin2018holistic,lin2016towards}, compact networks \cite{sandler2018mobilenetv2:,ma2018shufflenet}, knowledge distillation (KD) \cite{hinton2015distilling,romero2014fitnets}, neural architecture search (NAS) \cite{zoph2017neural,zheng2019multinomial} and quantization \cite{courbariaux2016binarized,jacob2018quantization}. Considering the unique structures such as EDSR \cite{lim2017enhanced} and RDN \cite{zhang2018residual} in SR, it is by nature to leverage quantization schemes to accelerate and compress SR networks, \emph{i.e.},  by converting full-precision weights \cite{jacob2018quantization}, activations \cite{cai2017deep}, and gradients \cite{zhou2016dorefa-net:} to low bits.

Ma \emph{et al.} \cite{ma2018efficient} first apply weight quantization to compress SR models, which merits in low on-device storage. However, the computational complexity is still significantly high, since full-precision activations are still used. In contrast, directly applying weight quantization to activations will incur significant accuracy drop in general SR tasks without using batch normalization, due to the high dynamic quantization range. On one hand, the work in \cite{lim2017enhanced} has shown that normalizing features on SR models limits the network's representation power. Since BN layers make the features to be smooth, which results in the blurred reconstructed HR images with artifacts. To this end, recent SOTA SR models (\emph{e.g.} EDSR \cite{lim2017enhanced}, RDN \cite{zhang2018residual}) have already removed BN layers to obtain better reconstructed HR images. On the other hand, the absence of BN causes a severe dynamic range problem when quantizing the activations by using the SOTA quantization methods \cite{jacob2018quantization,choi2018pact:}. For example, the work \cite{jacob2018quantization} simply set the upper scale of activations to their max value, which causes significant performance degeneration in SR task. This is due to the fixed max scale that may be an outlier as the upper scale. Although Choi \emph {et al.} \cite{choi2018pact:} propose PACT to clip and quantize activations by learnable parameters, it only concentrates on the positive range while neglecting the gradient information in the negative range. In addition, the novel regularization term  \cite{choi2018learning} is added to automatically learn quantized controlling parameters and then obtain an accurate low-precision model. However, it leads to the increase of additional computation burdens and memory footprint, which is not runtime friendly for practical applications.

To address the above issues, a novel quantization scheme, termed PArameterized Max Scale (PAMS), is proposed to compress and accelerate SR models. Different from the previous works that focus on quantizing activations in a fixed manner, PAMS adaptively explores the upper bound of quantization range based on the gradients using a trainable clamp function, which significantly improves the model generality. Furthermore, structured knowledge transfer (SKT) is introduced to transfer structured knowledge from the full-precision network to the quantized one, which enables the latter to gain better visual perception. Fig. \ref{framework} presents the flowchart of our method. We first replace each basic block in the SR model with PAMS block. In each PAMS block, weights are quantized before they are convolved with the inputs and activations are quantized after the outputs of convolutional layer with its own learnable max scale. To further improve the performance of the quantized model, we align the high-level features between full-precision model and the corresponding low-precision quantized one among pixels. Finally, we employ stochastic gradient descent (SGD) method to minimize the objective function, which leverages the distillation loss to pixel-wise loss. 
	
We evaluate our method on several benchmarks over widely-used deep SR models like EDSR \cite{lim2017enhanced} and RDN \cite{zhang2018residual}. Quantitative and qualitative results demonstrate that PAMS can well quantize various SR models with a significantly high compression ratio, as well as nearly identical accuracy to the full-precision SR models. The proposed PAMS also well outperforms most existing alternatives such as Dorefa \cite{zhou2016dorefa-net:}, Tensorflow Lite \cite{jacob2018quantization} and PACT \cite{choi2018pact:}. For instance, on BSD100, the 4-bit PAMS-EDSR outperforms 4-bit Dorefa-EDSR by 0.828dB with a scale factor of  $\times 4$. Extended experiments also show that SKT is more effective for the quantized SR models with lower-bit operations. 

\section{Related Work}

\noindent
	\textbf{Deep SR models with light weights.} 
	Most recent SR models are built based upon DCNNs, for instance, MDSR \cite{lim2017enhanced} and RDN \cite{zhang2018residual}. Such networks are typically deep with heavy computation cost and memory footprint, which restrict their applications in resource-limited devices. Recent advances in SR network compression mostly focus on redesigning light-weight networks. For instance, DRRN \cite{tai2017image} and DRCN \cite{kim2016deeply} have been proposed to share parameters for reducing network parameters. However, the cost of computation and memory storage in these networks are still very large, due to the floating-point operations during inference and the sufficient parameters to ensure the model capability. 
	
	\noindent
	\textbf{Network quantization.} Previous works in network quantization mainly focus on quantizing weights \cite{rastegari2016xnor}, while maintaining the full-precision activations to ensure the model performance. Joint quantization of activations and weights are explored in HWGQ \cite{cai2017deep} and PACT \cite{choi2018pact:}. However, these methods mainly concentrate on object classification \cite{krizhevsky2012imagenet,he2016deep}, which is easier than the complex pixel-wise or patch-wise SR tasks. The work in \cite{ma2018efficient} serves as the first to extend quantization to compress SR models, which quantizes only weights to be binary. However, the operations between activations and quantized weights are still floating-point, which cannot largely reduce the FLOPs towards the practical speedup. Different from the previous work \cite{ma2018efficient}, we optimize the SR network with both low-bit quantized weights and activations by introducing a learnable parameter, which achieves the bound of the quantization range.
		
	
	\noindent
	\textbf{Knowledge distillation.}
	Knowledge distillation \cite{hinton2015distilling} aims to transfer the knowledge from a cumbersome network (teacher) to a compact network (student). It has been widely applied to various computer vision tasks by using the softened output knowledge \cite{hinton2015distilling} and intermediate feature representations \cite{lin2018holistic,zagoruyko2017paying}. In line with our work, Zhuang \emph{et al.} \cite{zhuang2018towards} proposed a guidance loss to jointly optimize the full-precision network and the low-precision model. However, it is not suitable to directly use such probability-based loss for SR, as the outputs of SR are reconstructed HR images. Different from these methods, our approach adopts structured knowledge based on the implicit information of a pre-trained network, which concentrates on aligning the spatial correlation between the low-precision and full-precision features to be more suitable for pixel-wise SR task.


	\begin{figure}[!ht]
	\centering

	\setlength{\abovecaptionskip}{0cm}
	\setlength{\belowcaptionskip}{0cm}
	\includegraphics[width=0.96\columnwidth, height=0.5\textwidth]{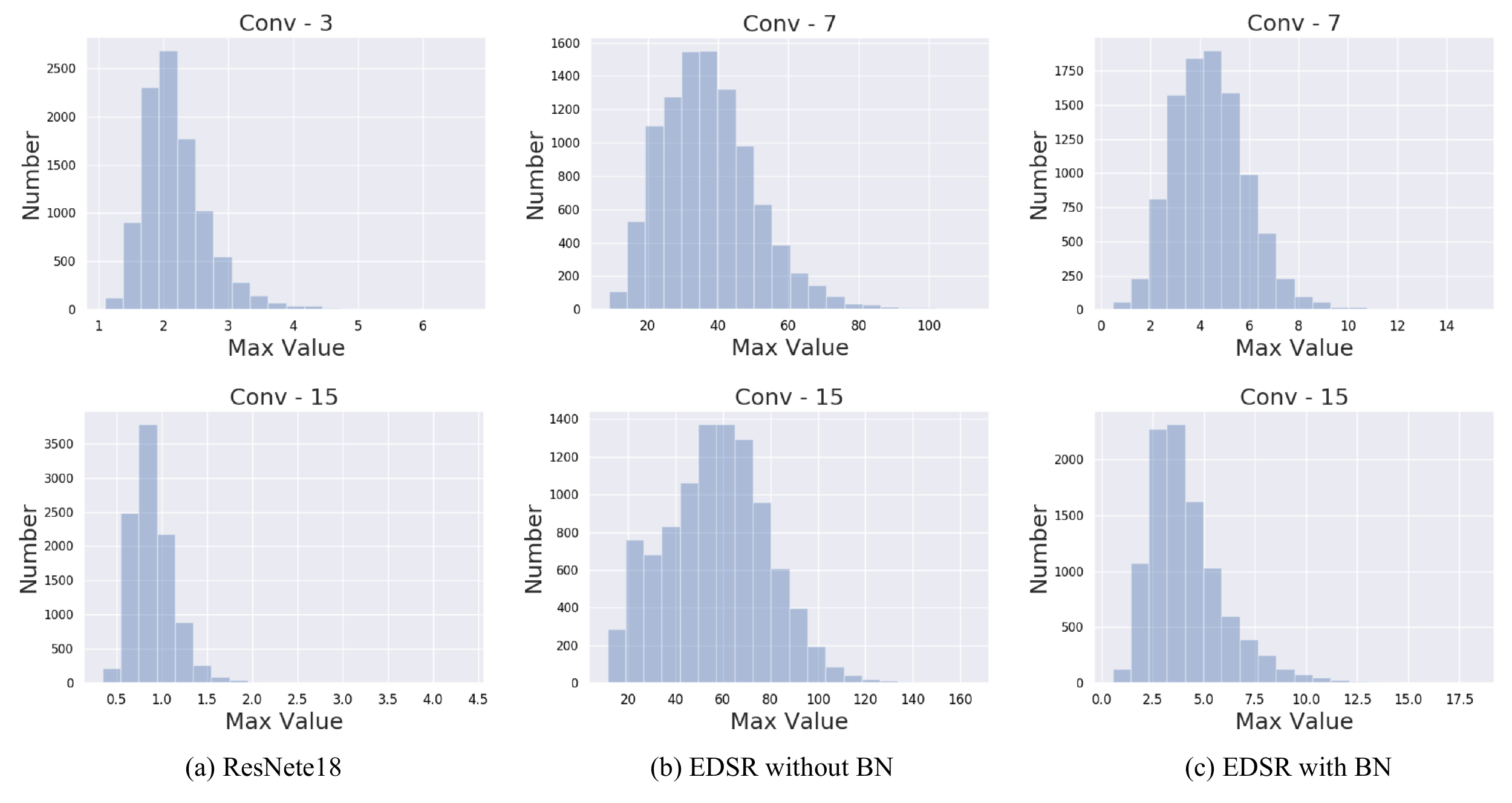} 
	
	\caption{Max value of activations from different layers and samples of ResNet-18 on ImageNet,  EDSR  \cite{lim2017enhanced} w. / wo.  BN on DIV2K. The absence of BN causes more dynamic range problem in SR models.} 
	\label{fig:kde}

	\end{figure}

	\section{The Proposed Method}	
	\subsection{A Close Look at SR Model Quantization}
	
	Current practice \cite{ma2018efficient} only quantizes the weights in deep SR models, which does reduce the storage cost but unfortunately ignores the computational efficiency caused by the full-precision multiplication between weights and activations. Moreover, the conversion between low-precision weights and full-precision activations aggravates the training time. It is not runtime-friendly to deploy such quantization scheme in real scenarios. Note that some quantization methods quantize the activations based on the premise of batch normalization \cite{ioffe2015batch}. In this way, the activations are supposed to stay in a stable range. 
	
	However, prior work \cite{lim2017enhanced} has shown that batch normalization layers get rid of range flexibility by normalizing the features, and simply removing them can make a big margin of improvements while reducing GPU memory cost. This modification can be effectively extended to the recent state-of-the-art SR models (\emph{e.g.} RDN \cite{zhang2018residual}, RCAN \cite{zhang2018image}, DBPN \cite{haris2018deep}) for ensuring range flexibility and reducing artifacts. Fig. {\ref{fig:kde}} shows the statistics collected from pre-trained ResNet-18 on ImageNet, EDSR with and without BN on DIV2K. We can see that the max value of activations varies a lot in different samples in the same layer, and the activation range is more dynamic of SR model (EDSR) than that in classification model (ResNet-18). It indicates the dynamic range problem is more severe in the SR model than that of the classification model. Moreover, the absence of BN causes a more severe dynamic range problem. Compared to EDSR with BN (Fig. \ref{fig:kde}(b)), the max value of activations in EDSR without BN has a wider max value range and shows a more even distribution, which indicates that removing BN in current SR models causes activations in a more dynamic range, so as to be difficult to manually decide the quantization range. We argue that this quantization range is vital to the performance: If its maximum tends to a tiny value, the upper bound of the quantization range will be very small. And as reflected in the reconstructed HR images, the details will be mostly lost, which thus causes significant quality degradation. In contrast, if its maximum is abnormally large, the quantization range may include outliers that contain redundant information and decrease the accuracy of quantized DNNs.

	\subsection{Parameterized Max Scale (PAMS)}
	The proposed PAMS quantizes both activations and weights of deep SR models. In this subsection, we first elaborate on our overall quantization approach. Then we describe how to leverage trainable truncated parameters to adaptively learn the upper bound of activations. To efficiently use the pre-trained network and improve the performance, we further introduce a structured knowledge transfer (SKT) loss. The overall framework is presented in Fig. \ref{framework}.
	
	\noindent
	\textbf{Quantization function.}
	As shown in \cite{cai2017deep}, distributions on different layers of activations tend to be symmetric. This characteristic can help to improve the model accuracy of a quantized network with extremely low-bit weights and activations, as validated in \cite{faraone2018syq:}. Therefore, given a specific full-precision model with a parameter set $\mathcal{X}$ ($\mathcal{X}$ denotes either weights or activations of a specific layer), we quantize every element $x$ ($x \in \mathcal{X}$) by using the following point-wise quantization function $Q$ with a symmetric mode: 
	\begin{equation} \label{quantization function}
	\begin{aligned}
	Q(x,n) = \lfloor \frac{f(x)}{s(n)} \rceil s(n),
	\end{aligned}
	\end{equation}
	where $f(x)$ is the clamp function that limits the inputs range and $s(n)$ is the map function that scales the higher precision inputs to their lower bit reflections, which can be formulated by $f(x) = max(min(x,a),-a)$ and $s(n) = \frac{a}{2^{n-1}-1}$, respectively. $n$ denotes the number of quantization level. $a$ represents the maximum of the absolute value of $\mathcal{X}$ and $\lfloor \cdot \rceil$ rounds the value to the nearest integer.
	
	
	For quantizing weights, previous work \cite{jacob2018quantization} has shown that simply set $a:=max(|w|)$ only has a negligible effect on the performance, which is therefore adopted in our approach. As for activations, the quantization range depends on the inputs, which leads to a dynamic range. This instability is unfavorable to the performance and model generality, which is designed in the following.


	\noindent
	\textbf{Trainable upper bound.}
	In the previous work \cite{choi2018pact:}, the dynamic range of activations can be partially alleviated by using a parameterized clipping activation function to replace the Rectified Linear Unit (ReLU), which limits the application scope.  In this paper, we propose a novel activation quantization scheme, in which the clamp function $f^*(\cdot)$ has the trainable parameter $\alpha$ to dynamically adjust the upper bound of the quantization range. We can directly employ the stochastic gradient descent to update this parameter, which is able to minimize the performance degradation arising from the quantization. For a given activation, the corresponding value will be quantized to $n$ bits by:
	\begin{equation} \label{pams scaling factor}
	\begin{aligned}
	x_q = Q(x,n) = \frac{\alpha}{2^{n-1}-1} \times \lfloor \tilde{x} \times \frac{2^{n-1}-1}{\alpha} \rceil,
	\end{aligned}
	\end{equation}
	where $\tilde{x} = f^{*}(x) = max(min(x,\alpha),-\alpha)$. The dynamic range is limited to $[-\alpha,\alpha]$. The advantages of our quantization function lie in that $Q(x,n)$ directly involves in the back-propagation process based on only one learnable parameter $\alpha$, based upon which we can train the quantized SR network in an end-to-end manner. Extensive experiments in Section \ref{experiments} demonstrate that Eq.\ref{pams scaling factor} can introduce more effectiveness comparing to several state-of-the-art quantization methods \cite{zhou2016dorefa-net:,jacob2018quantization,choi2018pact:}.
	

	\noindent
	\textbf{Back-propagation with quantization.} In back-propagation, $ \frac{\partial x_q}{\partial \tilde{x}}$ can be approximated to $1$ based on the straight-through estimator (STE) \cite{courbariaux2016binarized}. Inspired by \cite{choi2018pact:}, the gradient of $\alpha$ is calculated as follows:

	\begin{equation} \label{gradient_alpha}
	\frac{\partial x_{q}}{\partial \alpha} \approx \frac{\partial x_{q}}{\partial \tilde{x}} \frac{\partial \tilde{x}}{\partial \alpha} = \left\{\begin{array}{ll}{-1,} & {x \in(-\infty, -\alpha]}, \\ {0,} & {x \in(-\alpha, \alpha)}, \\ {1,} & {x \in[\alpha,+\infty)}. \end{array}\right.
	\end{equation}
	
	Note that,  the work in \cite{choi2018pact:} cuts off the gradients in the regions satisfied with $x < 0$, while PAMS can adaptively adjust $\alpha$ based on the gradients in both $x \ge \alpha $ and $x \le \alpha$ areas. It is important for the post-training quantization, since the gradients of the pre-trained model tend to 0. In other words, PAMS can retain more gradient information for updating $\alpha$.


	\noindent
	\textbf{Initializing \bm{$\alpha$}}.
	To avoid gradients vanishing or exploding, non-convex optimization on DCNNs heavily depends on the initialization of parameters. Instead of manually designing the initial value of $\alpha$, initialization based on the statistics from a pre-trained network can achieve better performance. Therefore, we resort to task-related statistical methods based on the pre-trained network to calibrate the quantization error. In particular, given the $l$-th layer with $m$ input activations ${x_{1}^{(l,t)}, ..., x_m^{(l,t)}}$, $\alpha^{(l)}$ is calculated by the redefined exponential moving average (EMA) function at the start of training:
	\begin{equation} \label{ema_function}
	\alpha^{(l,t)} = \beta \cdot\alpha^{(l,t-1)} + (1 - \beta) \cdot avg(max(x_{1}^{(l,t)}),..,max(x_{m}^{(l,t)}),
	\end{equation}	
	
	\noindent where $t$ is the iteration number and $\beta$ denotes the smoothing parameter of EMA, which is set to be 0.9997. Specially, we set $\beta$ to 0 when $t$ is 0. 	
%

\subsection{Optimization}


\begin{algorithm}
	\renewcommand{\algorithmicrequire}{\textbf{Input:}}
	\renewcommand{\algorithmicensure}{\textbf{Output:}}
	\caption{Quantization SR Model}
	\label{process}
	\begin{algorithmic}[1]
		\REQUIRE Training dataset $D$, full-precision model $T$, quantization level $n$; 
		\ENSURE The quantized model $S$;	 
		\STATE Define the low-precision model $S$ by replacing convolution layers of $T$ by $n$-bit PAMS;		
		\STATE {Initialize $\alpha^{(l)}$ of each layer $l$ with Eq. \ref{ema_function};}
		
		\FOR{$i=1,...,N$ epoch} 
			\STATE {Forward pass by applying clamp function to weights and activations using Eq. \ref{quantization function} and Eq. \ref{pams scaling factor}};
			\STATE {Update all parameters in Eq. \ref{total_sr_loss} via SGD;}
		\ENDFOR
		\STATE \textbf{return} $S$;
		
	\end{algorithmic} 
 
\end{algorithm}

	\noindent
	\textbf{Pixel-wise loss.} 
	Given a training dataset $D = \{I_{LR}^{i},I_{HR}^{i}\}_{i=1}^{n}$ with $n$ LR input images and their corresponding HR counterparts, SR models are commonly optimized by minimizing the conventional pixel-wise $L_1$ loss between the output $I_{SR}$ and the ground truth image  $I_{HR}$:
	
	\begin{equation} 
	\label{pixel-wise_loss}
	L_{PIX} = \frac{1}{n} \sum\limits_{i=1}^n||I_{HR}^{i}-I_{SR}^{i}||_{1},
	\end{equation}
	
	\noindent
	where $||\cdot||_1$ denotes the $L_1$ norm.
	A better SR model needs to infer the high-frequency textures from a low-resolution input. However, it is hard to obtain by only using Eq. \ref{pixel-wise_loss} based on low-bit quantization, which is due to the accumulated quantization error.

	\noindent
	\textbf{Structured knowledge transfer (SKT).}
	Inspired by \cite{zagoruyko2017paying}, we consider that the full-precision model has learned high-level representation, which provides knowledge to the low-precision one about where it concentrates. More specifically, instead of using the soft probability in the classification task, we align the structured features between the cumbersome network and the quantized one by minimizing their pixel-wise distance. Therefore, the loss function for our SKT is defined as:
	
	\begin{equation} \label{gradient of alpha}
	L_{SKT} = || \frac{F'_{S}}{||F_{S}'||_{2}}-\frac{F_{T}'}{||F_{T}'||_{2}}||_{p},
	\end{equation}
	
	\noindent
	where $F_{T}'$, $F_{S}'$ are a pair of structure features after the spatial mapping of activations from the full-precision network and the correspond low-precision one, respectively. The spatial mapping defined by $F' = \sum_{i=1}^{C}\left|F_{i}\right|^2 \in \mathbb{R}^{H \times W}$, where $F \in \mathbb{R} ^{C \times H \times W}$ denotes the activations after the last layer in the high-level feature extractor. We set $p=2$ for $p$-norm in our experiments. In sum, SKT enhances the learning process of spatial correlation in the low-precision model which effectively improves the performance of the quantized network and provides an additional constraint to avoid producing over-smoothed images.

\setlength{\tabcolsep}{4pt}
\begin{table}[!ht]
\begin{center}

\caption{Comparison between quantizing EDSR \cite{lim2017enhanced} and RDN \cite{zhang2018residual} by deploying PAMS with low-bit weights and activations on the public benchmark (PSNR(dB)/SSIM). The higher PSNR and higher SSIM, the better performance the methods achieve. EDSR is based on the residual block and RDN is based on the dense block. RDN* denotes the results based on our implementation.} 

\label{edsr_rdn_quantization}
\resizebox{\textwidth}{18mm}{
\begin{tabular}{ccc|ccc|cccc}

\toprule
\noalign{\smallskip}

Dataset & Scale & Bicubic & EDSR & \tabincell{c}{PAMS-EDSR\\(8-bit)} & \tabincell{c}{PAMS-EDSR\\(4-bit)} & RDN* & \tabincell{c}{PAMS-RDN \\(8-bit)} & \tabincell{c}{PAMS-RDN \\(4-bit)} \\  
\noalign{\smallskip}

\hline
\noalign{\smallskip}

\multirow{2}*{Set5} & $\times$2 & 33.66/0.9299 & 37.985/0.9604 & 37.946/0.9603 & 37.665/0.9588 & 38.027/0.9606 & 38.060/0.9606 & 36.528/0.9527  \\
 & $\times$4 & 28.42/0.8104 & 32.095/0.8938 & 32.124/0.8940 & 31.591/0.8851& 32.244/0.8959 & 32.340/0.8966 & 30.441/0.8624 \\ 

\noalign{\smallskip}
\hline
\noalign{\smallskip}

\multirow{2}*{Set14} &  $\times$2 &30.24/0.8688  & 33.568/0.9175 & 33.564/0.9175 & 33.196/0.9146 & 33.604/0.9174 & 33.732/0.9189 & 32.392/0.9050 \\
& $\times$4 & 26.00/0.7027 & 28.576/0.7813 &   28.585/0.7811 & 28.199/0.7725 & 28.669/0.7838 & 28.721/0.7848 & 27.536/0.7530 \\

\noalign{\smallskip}
\hline
\noalign{\smallskip}

\multirow{2}*{BSD100}  & $\times$2  & 29.56/0.8431 & 32.155/0.8993 & 32.157/0.8994 & 31.936/0.8966 & 32.187/0.8999 & 32.215/0.9000 & 31.268/0.8853 \\
& $\times$4 &  25.96/0.6675 & 27.562/0.7355 & 27.565/0.7352 & 27.322/0.7282  & 27.627/0.7379 & 27.644/0.7382 &  26.869/0.7097 \\

\noalign{\smallskip}
\hline
\noalign{\smallskip}

\multirow{2}*{Urban100}  & $\times$2  & 26.88/0.8403  & 31.977/0.9272 & 32.003/0.9274 & 31.100/0.9194 & 32.084/0.9284 &   32.262/0.9298  & 29.703/0.8976  \\
& $\times$4 &  23.14/0.6577 & 26.035/0.7848 & 26.016/0.7843 & 25.321/0.7624 &  26.293/0.7924 & 26.367/0.7955 & 24.523/0.7256 \\

\noalign{\smallskip}
\bottomrule
\end{tabular}
}
\end{center}
\end{table}

	\noindent	
	\textbf{The overall loss function.}
	Given an SR model, as consistent with the distillation term mentioned above, the whole objective function is given as:
	
	\begin{equation} \label{total_sr_loss}
	L_{SR} = \lambda_{p} L_{PIX} +\lambda_{s} L_{SKT},
	\end{equation}
	where $\lambda_{p}$ and $\lambda_{s}$ are coefficients to control the balance of the corresponding loss. We set $\lambda_{p}$ to 1 and $\lambda_{s}$ to $10^3$. The overall optimized process is summarized in Alg. \ref{process}.

	\section{Experiments \label{experiments}}
	\subsection{Experimental Settings}

	\noindent
	\textbf{Datasets and metrics.}
	DIV2K \cite{timofte2017ntire} contains 800 training images, 100 validating images and 100 testing images. We train all models with DIV2K training images. For testing, we use four standard benchmark datasets: Set5 \cite{bevilacqua2012low-complexity}, Set14 \cite{ledig2017photo-realistic}, BSD100 \cite{martin2001a} and Urban100 \cite{huang2015single}. For the evaluation metrics, we use PSNR and SSIM \cite{wang2004image} over the Y channel between the output quality image and the original HR image.

		\noindent
	\textbf{SR models and alternative approaches.} 
	Both residual block and dense block are widely used in SR models, like VDSR \cite{kim2016accurate}, EDSR \cite{lim2017enhanced} and RDN \cite{zhang2018residual}. To validate the superiority of our approach, we choose EDSR and RDN as backbones and use 8-bit and 4-bit quantization on them. As most parameters exist in the high-level feature extraction module, we do not quantize weights and activations in low-level feature extraction and reconstruction modules, which ensures a trade-off between performance and model size. The qualitative comparisons are generated by the publicly available source code in EDSR \cite{lim2017enhanced}.


\begin{figure*}[!ht]
\begin{center}

\includegraphics[width=0.99\linewidth, height=0.98\textwidth]{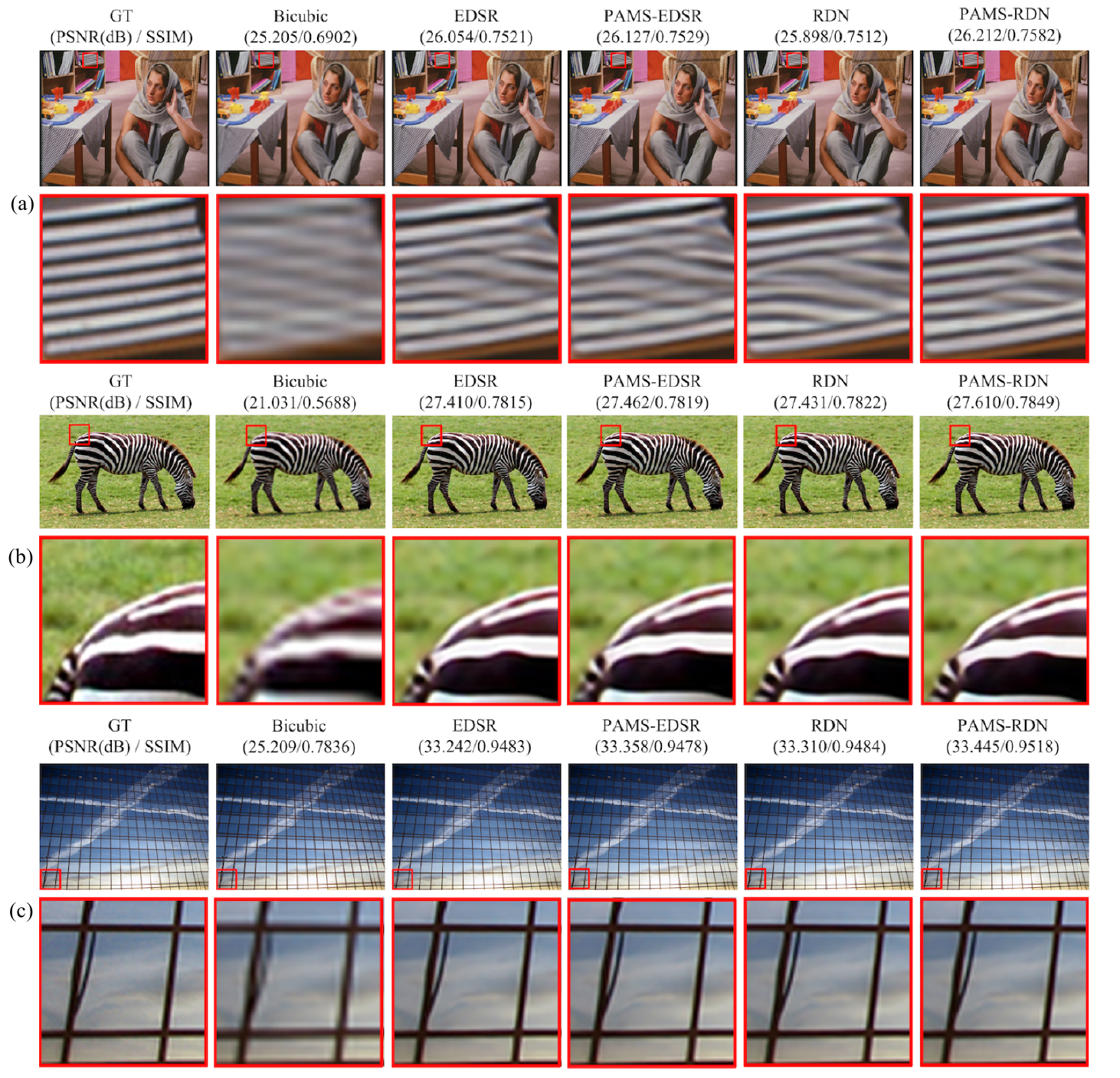}
\setlength{\abovecaptionskip}{-1cm}
\end{center}

   \caption{Qualitative comparison between 8-bit and full-precision models with a scale factor of $\times 4$. (a) and (b) are the results of ``barbara'' and ``zebra'' from Set14, respectively. (c) is the results of ``img055'' from Urban100. Note that the quantized models with PAMS produce extremely similar or even better SR images to their full-precision counterparts, while the former has a significant reduction of model size and computational complexity.}
\label{fig:exps}

\end{figure*}

	\noindent
	\textbf{Training setting.}
	The model is implemented by using PyTorch \cite{paszke2017automatic}. Following the setting of \cite{lim2017enhanced}, we pre-process all images in the DIV2K training dataset by subtracting the mean RGB and adopt a normal data augmentation during training, which includes random horizontal flips and vertical rotations. The mini-batch size is set to 16. We deploy the ADAM optimizer with $\beta_1 = 0.9$, $\beta_2 = 0.999$ and $\epsilon = 10^{-8}$ to the model, which is trained for 30 epochs. The learning rate is initialized by $10^{-4}$ and is halved at every 10 epochs.

\setlength{\tabcolsep}{8pt}

\begin{table}[!ht]

\begin{center}

\caption{Comparison to the state-of-the-art quantization methods by using different bits on a scale factor of $\times 4$ super-resolution. EDSR is the backbone network.}\smallskip
\label{tab:quant_compare}

\resizebox{\textwidth}{21mm}{
\begin{tabular}{cccccc}
\toprule
\noalign{\smallskip}
Dataset & Bits & Dorefa-EDSR & TF Lite-EDSR & PACT-EDSR & PAMS-EDSR  \\  
\noalign{\smallskip}
\hline
\noalign{\smallskip}

Set5 & \tabincell{c}{8\\4}  & 
					\tabincell{c}{ 30.194/0.8556 \\ 29.569/0.8369 } & 
					\tabincell{c}{ 31.910/0.8906\\31.380/0.8812  }  & 
					\tabincell{c}{ 31.520/0.8853 \\ 31.393/0.8834 }  & 
					\tabincell{c}{ \textbf{32.124/0.8940} \\ \textbf{31.591/0.8851} } \\
\noalign{\smallskip}
\hline
\noalign{\smallskip}
Set14 & \tabincell{c}{8\\4}  & 
					\tabincell{c}{  27.297/0.7492 \\  26.817/0.7352 }& 
					\tabincell{c}{  28.416/0.7779 \\28.109/0.7690 }  &  
					\tabincell{c}{ 28.181/0.7712 \\ 28.104/0.7695 }  &  
					\tabincell{c}{  \textbf{28.585/0.7811} \\ \textbf{28.199/0.7725}	 } \\
\noalign{\smallskip}		
\hline
\noalign{\smallskip}
BSD100 & \tabincell{c}{8\\4}  &  
					\tabincell{c}{  26.767/0.7079 \\ 26.474/0.6971} & 
					\tabincell{c}{ 27.470/0.7329 \\ 27.252/0.7239 }  & 
					\tabincell{c}{ 27.288/0.7261 \\ 27.251/0.7245  }  & 	
					\tabincell{c}{ \textbf{27.565/0.7352} \\ \textbf{27.322/0.7282}	 } \\
\noalign{\smallskip}			
\hline
\noalign{\smallskip}
Urban100 & \tabincell{c}{8\\4}  & 
					\tabincell{c}{ 24.220/0.7128 \\ 23.753/0.6898 } & 
					\tabincell{c}{ 25.739/0.7760  \\ 25.198/0.7551 }  & 
					\tabincell{c}{ 25.245/0.7570 \\ 25.148/0.7535  }  & 
					\tabincell{c}{ \textbf{26.016/0.7843} \\ \textbf{25.321/0.7624}	 } \\			
\noalign{\smallskip}
\bottomrule
\noalign{\smallskip}
\end{tabular}
}

\end{center}
\end{table}


\begin{figure*}[ht]
	\centering

	\setlength{\abovecaptionskip}{0cm}
	\setlength{\belowcaptionskip}{0cm}
	\includegraphics[width=1.0\textwidth, height=0.43\textwidth]{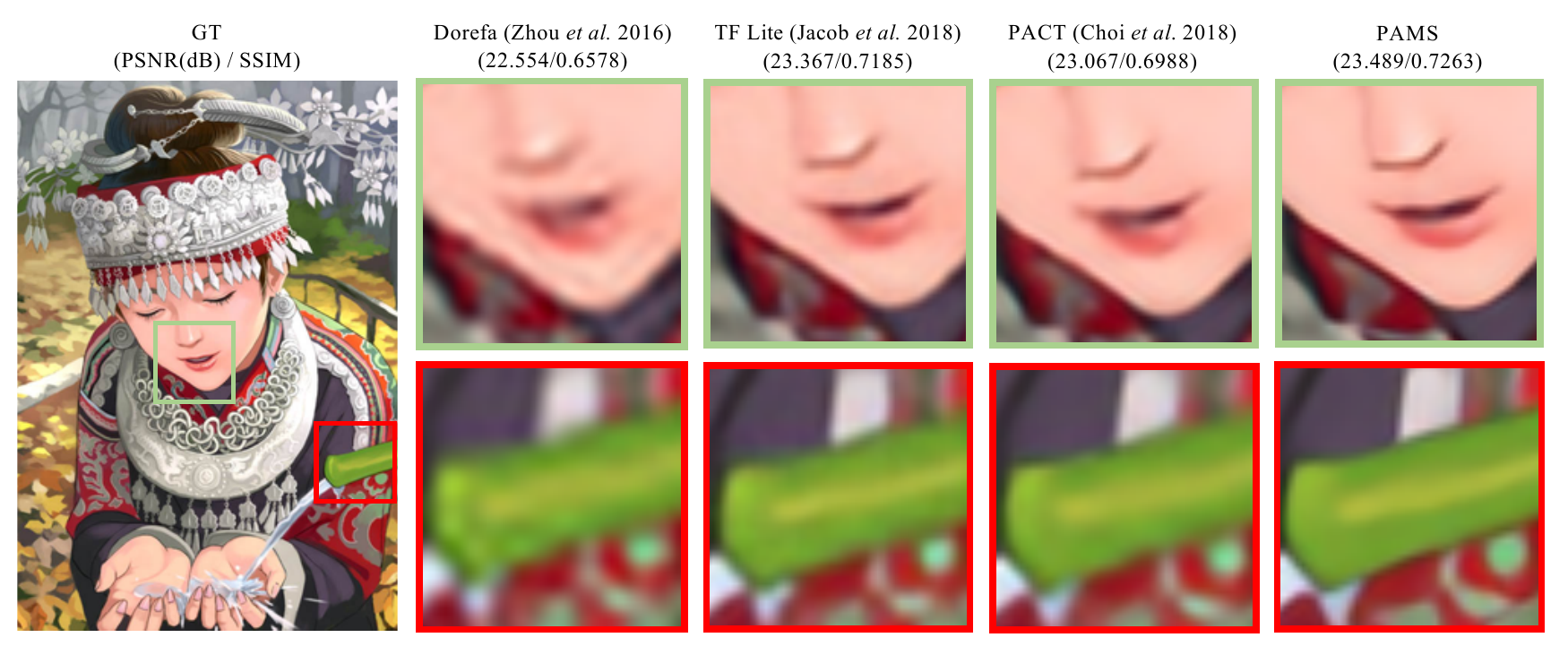} 
	\caption{Qualitative comparison of our method with other quantization methods on a scale factor of  $\times 4$.}
	\label{fig:quant_methods}

\end{figure*}


	\subsection{Quantitative and Qualitative Results}
	As shown in Table. \ref{edsr_rdn_quantization}, the proposed PAMS with 8-bit weights and activations achieves competitive or even better results on different backbones. For instance, 8-bit PAMS-RDN outperforms the full-precision RDN by $0.178$dB PSNR and $0.074$dB PSNR on Urban100 with scale factors of $\times 2$ and $\times 4$, respectively. The 4-bit PAMS-EDSR only suffers $0.24$dB PSNR loss on BSD100 for a scale factor of  $\times 4$ compared to its full-precision model.  Quantizing RDN leads to a significant improvement over EDSR in 8-bit, which indicates that dense blocks may produce more redundancy than residual blocks. We provide more qualitative evaluations on the 8-bit quantization in Fig. \ref{fig:exps}. The models with PAMS produce more visually natural images than the bicubic interpolation, and are extremely similar to their full-precision counterparts. Considering that residual-based models are widely used, the results also indicate the generality of the proposed method.



\setlength{\tabcolsep}{4pt}
\begin{table}
\begin{center}
\caption{Comparison of EDSR and RDN with different bits on BSD100. W and A represent the number of bits of weights and activations, respectively.} \smallskip
\label{tab:compress_ratio_edsr}
\begin{tabular}{ccccc}

\toprule\noalign{\smallskip}
Model & W/A  & StorageSize ($r_{comp}$) & PSNR(dB) / SSIM  \\
\noalign{\smallskip}
\hline
\noalign{\smallskip}

EDSR (32-bit)& 32/32 &  1.518M (0\%)	& 27.562/0.7355\\

PAMS-EDSR (8-bit)& 8/8	&  0.631M (58.4\%)&27.565/0.7352 \\
PAMS-EDSR (4-bit)& 4/4	& 0.484M (68.1\%) &27.322/0.7282 \\
\noalign{\smallskip}
\hline
\noalign{\smallskip}
RDN (32-bit)& 32/32 &   22.27M (0\%)	&27.627/0.7379 \\
PAMS-RDN (8-bit)& 8/8	&  5.82M (73.9\%)&27.644/0.7382 \\
PAMS-RDN (4-bit)& 4/4	& 3.08M (86.2\%) &26.869/0.7097 \\
\noalign{\smallskip}
\bottomrule
\noalign{\smallskip}
\end{tabular}
\end{center}

\end{table}

For a better comparison, we re-implement Dorefa \cite{zhou2016dorefa-net:}, Tensorflow Lite \cite{jacob2018quantization} and PACT \cite{choi2018pact:} on EDSR. We use the same initialization method and quantize both weights and activations in each residual block as PAMS-EDSR. For Dorefa, we do not quantize gradients for a fair comparison. Table. \ref{tab:quant_compare} shows the results of 8-bit and 4-bit EDSR. Our method achieves better performance, compared to all baselines. For example, 8-bit PAMS-EDSR outperforms 8-bit Dorefa-EDSR by 1.288dB PSNR and 1.796dB PSNR on Set14 and Urban100, respectively. The reconstruction results are further shown in Fig. \ref{fig:quant_methods}. Compared to other methods. The output (SR images) using PAMS are better-looking with sharp edges and rich details. In conclusion, PAMS with trainable truncated parameters rely on the backward which achieves much better generalization ability.


 \begin{figure*}[!ht]
	\centering
	\includegraphics[width=0.96\textwidth, height=0.17\textwidth]{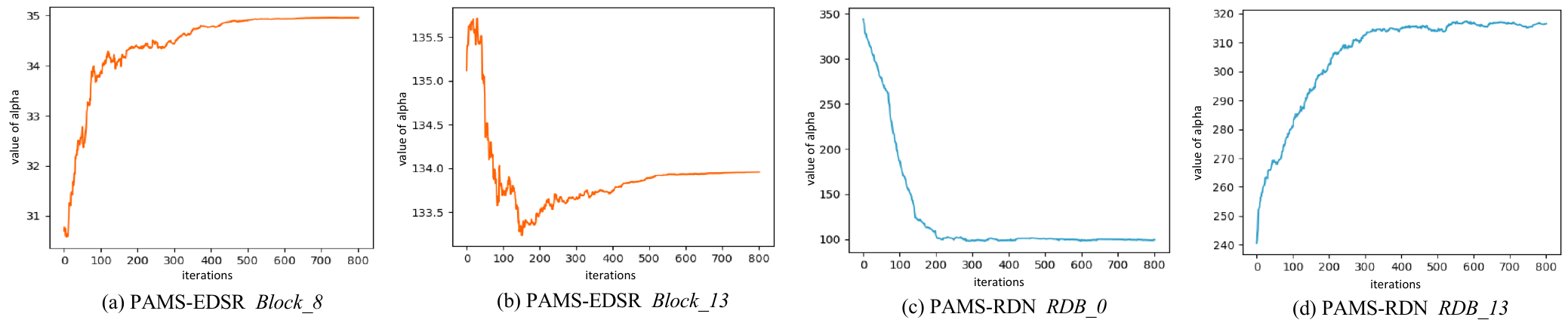} 

	\caption{Convergence curves of $\alpha$ for 8-bit PAMS-EDSR and 8-bit PAMS-RDN.}
	\label{fig:act}
\end{figure*}


\subsection{Compression Ratio}
 The model size and compression ratio of EDSR and RDN are presented in Table. \ref{tab:compress_ratio_edsr}. In particular, the full-precision network is represented by using single precision floating point. The model size of the full-precision network considerably decreases after quantization. Note that, we only quantize the weights and activations in the high-level feature extractor module, such that the compression ratios are calculated based on the total parameters of the network and the parameters in the high-level feature extractor. Although PAMS introduces a trainable parameter $\alpha$, it still yields a $50$\%-$90$\% compression ratio, since it directly depends on the backbone and the number of bits. It can be seen that 4-bit weights and activations cause more performance degradation than the 8-bit model. But lower-bit quantized networks can significantly reduce storage requirement.

\subsection{Convergence of the $\alpha$}
To demonstrate the convergence of our method, we directly validate the convergence on $\alpha$ during training. The results are presented in Fig. \ref{fig:act}. The first and second columns show the $\alpha$ of PAMS-EDSR on the layer of $Block\_8$ and $Block\_13$, respectively. The third and fourth columns show the $\alpha$ of PAMS-RDN in $RDB\_0$ and $RDB\_13$, respectively ($RDB$ denotes the Residual Dense Block). It illustrates that $\alpha$ in different layers not only have different values but also have different evolving directions. 


\setlength{\tabcolsep}{3pt}
\begin{table}[!ht]

\begin{center}

\caption{Comparison of the performance gap between the singe-precision EDSR, 8-bit PACT-EDSR w. / wo. BN and 8-bit PAMS wo. BN. (PSNR(dB) / SSIM).}
\label{tab:with_bn}
\begin{tabular}{cccccc}

\toprule
\noalign{\smallskip}
Model & With BN & Set5 & Set14 & BSD100 & Urban100 \\
\noalign{\smallskip}
\hline
\noalign{\smallskip}

PACT-EDSR & \checkmark &  0.531/0.0083	&  0.273/0.0068 & 0.166/0.0056 & 0.354/0.0125 \\
PACT-EDSR & $\times$ & 0.575/0.0085 &  0.395/0.0101 & 0.274/0.0094 & 0.790/0.0278 \\
PAMS-EDSR & $\times$  &  0.029/0.0002 &  0.009/-0.0002 & 0.003/-0.0003 & -0.019/-0.0005 \\

\bottomrule
\end{tabular}
\end{center}
\end{table}



\setlength{\tabcolsep}{4pt}
\begin{table}[!ht]

\begin{center}

\caption{Results about different initialization methods of $\alpha$ on EDSR with $\times 4$ scale factor (PSNR(dB) / SSIM).}
\label{tab:init_alpha}
\begin{tabular}{ccccc}

\toprule
\noalign{\smallskip}
Init. & Set5 & Set14 & BSD100 & Urban100 \\
\noalign{\smallskip}
\hline
\noalign{\smallskip}

Random &  31.782/0.8896  &  28.383/0.7779  & 26.273/0.6879  & 23.488/0.6780  \\
EMA & 32.002/0.8923 &  28.497/0.7797 & 28.497/0.7797 & 25.806/0.7788 \\

\bottomrule
\end{tabular}
\end{center}
\end{table}


For instance, PAMS-EDSR $Block\_8$ (Fig. \ref{fig:act}(a)) and PAMS-RDN $RDB\_0$ (Fig. \ref{fig:act}(c)) act in the same direction, while PAMS-EDSR $Block\_13$ (Fig. \ref{fig:act}(b)) and PAMS-RDN $RDB\_13$ (Fig. \ref{fig:act}(d)) are with the opposite trend. We also found that $\alpha$ can promote the convergence to a stable value for both EDSR and RDN, which indicates the effectiveness of our method.

\subsection{Ablation Study}  

\noindent
\textbf{Effect of BN in SR models.}
To investigate the effect of quantizing normalized features, we use PACT to quantize EDSR with BN and without BN. As shown in Table. \ref{tab:with_bn}, the performance gap between the quantized EDSR without BN is larger than the quantized EDSR with BN. For example, The gap of 8-bit PACT-EDSR without BN is 0.790dB PSNR on Urban100, which is larger than PACT-EDSR with BN (0.354dB PSNR). It shows that the performance degradation of unnormalized features is more pronounced in lower-precision SR models, Moreover, PAMS-EDSR can save more important information for unnormalized weights and activations which largely decrease the performance gaps.

%
%
%
%
%
%
%
%
%
%
%


\setlength{\tabcolsep}{4pt}
\begin{table}[!ht]
\begin{center}
\caption{Results of PAMS-EDSR w. / wo. $L_{SKT}$ on 8-bit and 4-bit settings (PSNR(dB) / SSIM).}
\label{tab:at_loss}
\begin{tabular}{ccccc}

\toprule\noalign{\smallskip}
Dataset & bits & without $L_{SKT}$ & with  $L_{SKT}$ & metrics. $\uparrow$ \\
\noalign{\smallskip}
\hline
\noalign{\smallskip}

\multirow{2}*{Set5} & 8 &32.127/0.8939& 32.124/0.8940 & -0.003/0.0001 \\
& 4  & 31.538/0.8842 & 31.591/0.8851 & \textbf{0.053/0.0009}\\

\noalign{\smallskip}
\hline
\noalign{\smallskip}

\multirow{2}*{Set14} & 8 &28.541/0.7807 & 28.585/0.7811 &  \textbf{0.044/0.0004}  \\
& 4  & 28.177/0.7723 &  28.199/0.7725 & 0.022/0.0002	 \\

\noalign{\smallskip}
\hline
\noalign{\smallskip}
\multirow{2}*{BSD100} & 8 & 27.550/0.7352 & 27.565/0.7352& 0.015 /0.0000 \\
& 4  & 27.302/0.7280 & 27.322/0.7282 & \textbf{0.020/0.0002}\\

\noalign{\smallskip}
\hline
\noalign{\smallskip}
\multirow{2}*{Urban100} & 8 & 25.984/0.7835 & 26.016/0.7843 & 0.032/0.0008 \\
& 4  & 25.250/0.7607 & 25.321/0.7624 &  \textbf{0.071/0.0017	} \\

\bottomrule
\end{tabular}
\end{center}
\end{table}


\noindent
\textbf{Effect of the learnable $\alpha$.} 
We compare our learnable max scale (PAMS) with the fixed maximum (TF Lite) for quantizing activations. Quantitative and qualitative results are represented in Table. \ref{tab:quant_compare} and Fig. \ref{fig:quant_methods}, respectively. Compared to TF Lite-EDSR, PAMS-EDSR achieves a better score as it produces sharper images and more realistic textures. It indicates that our method can learn a more suitable quantization range which contains more information about the full-precision model and reduces the quantization error.

\noindent
\textbf{Effect of the initialization of $\alpha$.} 
We evaluate our EMA initialization with random initialization on EDSR with a scale factor of $\times 4$. For the random mode, we initialize $\alpha$ in the activation quantization layer with a random number ranges from 0 to 128, which ensures that $\alpha$ can be initialized to a larger value in different layers independently. As illustrated in Table. \ref{tab:init_alpha}, EMA initialization achieves better performance on all benchmark datasets. To explain, EMA achieves better statistical distribution by $\alpha$ that can further help improve SR performance.

\noindent
\textbf{Investigating SKT loss.}
 To investigate the effectiveness of SKT, we further compare the quantized model with and without SKT. As shown in Table. \ref{tab:at_loss}, PAMS-EDSR which is optimized with the SKT outperforms the corresponding counterpart. Especially, our method obtains much better performance on lower bits. For instance, compared to the PAMS-EDSR without $L_{SKT}$ on Urban100, 4-bit PAMS-EDSR with $L_{SKT}$ gains $0.071$dB PSNR while 8-bit PAMS-EDSR with the same optimization gains only $0.032$dB PSNR. It also indicates that the feature maps from the full-precision model can help the low-precision model to better capture the spatial correlation from images.

\section{Conclusion}
In this paper, we propose a novel symmetric quantization scheme, termed PArameterized Max Scale (PAMS), to effectively quantize both weights and activations of the full-precision network for SR tasks. The proposed method adopts a truncated parameter $\alpha$ to adaptively adjust the upper bound of quantization range. This technique alleviates the  negative effect of dynamic range caused by the absence of batch normalization layers and helps to reduce the quantization error. To further approximate the full-precision network, we employ structured knowledge transfer (SKT) to retrain the quantized network in a few epochs. We have comprehensively evaluated the performance of the proposed approach on EDSR and RDN over public benchmarks, which demonstrates the superior performance gains and significant reduction in model size and computational complexity.

\noindent
\textbf{Acknowledgements.} This work is supported by the Nature Science Foundation of China (No.U1705262, No.61772443, No.61572410, No.61802324 and No.61702136), National Key R\&D Program (No.2017YFC0113000, and No.2016Y
FB1001503), Key R\&D Program of Jiangxi Province (No.20171ACH80022) and Natural Science Foundation of Guangdong Provice in China (No.2019B1515120049).

%
%
\bibliographystyle{splncs04}
\bibliography{egbib}
\end{document}